\documentclass[prl,twocolumn,showpacs,superscriptaddress]{revtex4}
\usepackage{amsmath,graphicx}

\begin{document}
\title{Granular Elasticity without the Coulomb Condition}

\author{Yimin Jiang}\email{yimin.jiang@uni-tuebingen.de}
\affiliation{Theoretische Physik, Universit\"{a}t T\"{u}bingen,
72076 T\"{u}bingen, Germany} \affiliation{Applied Physics and
Heat-Engineering, Central South University, Changsha
410083, China}
\author{Mario Liu}\email{mliu@uni-tuebingen.de}
\affiliation{Theoretische Physik, Universit\"{a}t T\"{u}bingen,
72076 T\"{u}bingen, Germany}
\date{\today}

\begin{abstract}
An self-contained elastic theory is derived which
accounts both for mechanical yield and shear-induced
volume dilatancy. Its two essential ingredients are
thermodynamic instability and the dependence of the
elastic moduli on compression.
\end{abstract}  \maketitle

Stable sand piles show clear elastic behavior, though
conventional elasticity theory cannot be appropriate, as
sand piles also possess a steepest slope. The associated
angle, the Coulomb angle $\varphi$, has a typical value
of around $30^\circ$ for dry sand. So granular materials
may be taken to interpolate between fluid (no elasticity,
$\varphi=0^\circ$) and solid (conventional elasticity,
$\varphi\ge90^\circ$). This is usually understood in
terms of the solid friction law: Subject to gravitational
pull, a sand grain resting on a slope of angle $ \theta$
experiences the force $N=\rho G\cos\theta$ normal to the
surface, and $S=\rho G\sin\theta$ along it, with
$S/N=\tan\theta$. ($\rho$ is the mass density and $G$ the
gravitational constant.) Since static friction will only
sustain a maximal $S/N$, there is a maximal $ \theta $,
which one may identify with the Coulomb angle $\varphi$,
or $S/N\leq\tan\varphi$. Now, requiring this to hold both
along any plane and everywhere in the bulk, one may
reinterpret $S$ and $N$ as components of the stress
tensor $\sigma_{ij}$, respectively tangential and normal
to the given plane. Then the inequality may, with
$\sigma_1$ and $\sigma_3$ as the largest and smallest
Eigenvalues of $\sigma_{ij}$, be written as
\begin{equation}
\left| (\sigma _1-\sigma _3)/(\sigma _1+\sigma _3)\right|
\leq \sin \varphi. \label{cy}
\end{equation}
This is the ``{\it Coulomb yield condition,}" or ``{\it
Coulomb law of internal friction}," a textbook formula of
soil mechanics~\cite{Nedderman}, employed to impose
mechanical yield upon conventional elasticity. Although
this formula captures essential granular physics, it
possesses a number of deficiencies. First of all, the
Coulomb condition ``pre-empts" what should have been the
result of a proper theory: Ideally, one would like to
start from a continuum theory, calculate the spatial
dependence of the stress with appropriate boundary
conditions, and arrive -- with some substantiated
understanding -- at the fact that there is a maximal
shear stress in sand, above which the system is
mechanically unstable. This includes especially an
expression for the Coulomb angle $\varphi$. The Coulomb
law postulates the last bit and employs it backward.

Second, the Coulomb condition takes for granted that
mechanical yield is determined by a unique $\varphi$,
independent of geometry. Third, it is not obvious that
granular materials behave as conventional solids up to
the yield point, without any ``precursor" behavior.
Fourth, the Coulomb condition partially contradicts
conventional elasticity, and our understanding is
rendered regrettably precarious. Last not least, Reynolds
dilatancy~\cite{Reynolds} -- the volume expansion
concurring with shear motion in granular materials --
should be an integral part of mechanical yield, yet is
completely ignored by the Coulomb law. Imagine a pile of
stacked steel balls, and envisage how a shearing
displacement lift the balls from their close-packed
positions and give rise to volume expansion -- which
eventually leads to yield. In experiments~\cite{Bagnold}
and simulations~\cite{Herrmann}, this is what has been
observed.

Although the Coulomb condition appears unique to granular
systems, its sole purpose is to account for mechanical
yield, or the lack of elastic configurations for certain
values of stress and strain. Yield is a widespread
phenomenon in many solids at high stresses, which are
well accounted for by linear elasticity at lower values
of stress. ``Low, high" are of course relative concepts,
and the noteworthy point is: Being characterized by a
quadratic elastic free energy and a linear stress,
\begin{eqnarray}\label{l-f}
f_{el}=\textstyle{\frac 12} K_bu_{nn}^2+K_au_{k\ell
}^0u_{k\ell }^0,\\ \sigma _{ij} =-K_bu_{nn}\delta _{ij}-
2K_au_{ij}^0, \label{l-str}
\end{eqnarray}
linear elastic theory cannot possibly account for yield,
as it provides stable elastic solutions for arbitrary
strains and stresses, however high. ($u_{nn}$ is the
trace of the strain tensor $u_{ij}$, $u_{kl}^0\equiv
u_{kl}-u_{nn}\delta _{kl}/3$ is its traceless part, and
$K_b,K_a$ are the constant compressional and shear
moduli.) Yet contrary to prevalent perception, this is
not a general feature of elasticity: Adding nonlinear
terms to Eqs~(\ref{l-f},\ref{l-str}) may well render
given elastic solutions instable for some variable range.
And it appears obvious that elastic instabilities, or
more generally, the lack of elastic solutions, are to be
identified with yield, the lack of elastic
configurations. Doing so embraces yield as the generic
phenomenon that it is, and does away with extraneous
inputs such as the Coulomb condition. Note also that
yield therefore marks the end of the range of validity
for elasticity. Only a more comprehensive theory
including dissipative terms is able to describe what then
happens, typically plastic flows.

In granular materials, one need not look far for terms
that fit the above description. Consider two solid
spheres in contact~\cite{LL7} to find $U\sim h^{5/2}$,
$f\sim h^{3/2}$, where $U$ is the elastic energy, $f$ the
applied force, and $h$ the relative change in height. The
latter relation is not linear, because the area of
contact between the two spheres and the amount of
compressed mass increase with $h$. Assuming one can scale
up this two-body result to granular materials -- in usual
parlance, that they possess {\em ``Hertz contacts,"} we
identify $U, f, h$ respectively with $f_{el},
\sigma_{nn}, u_{nn}$ of Eqs~(\ref{l-f},\ref{l-str}), and
conclude $K_b\sim{u_{nn}}^{1/2} \sim{\sigma_{nn}}^{1/3}$
is no longer constant. Realistically, with more than two
grains in contact, $K_b$ will still depend on
$\sigma_{nn}$ and vanish with it, so one may more
generally take $K_b\sim {\sigma_{nn}}^\beta$. And since
the physics of increasing contact area is similar for
shear, also take $K_a\sim {\sigma_{nn}}^\beta$. Evesque
and de Gennes employed these elastic moduli in the stress
of Eq~(\ref{l-str}) to successfully render the pressure
saturating in silos~\cite{EdG}.

Aiming to generalize this {\em ``quasi-elastic theory"}
and embed it into a consistent thermodynamic framework,
we made the following observation: With
$K_b,K_a\sim{u_{nn}}^a$ and the free energy $f_{el}$
retaining its form of Eq~(\ref{l-f}), the stress
$\sigma_{ij}$ -- given by general considerations
essentially as $\partial f_{el}/\partial u_{ij}$ -- is
necessarily modified: Eq~(\ref{l-str}) is clearly only
correct if $K_a,K_b$ are constant. Our serendipitous
finding reported below is, the additional terms of
$\sigma_{ij}$ suffice to account for mechanical yield and
volume dilatancy, by rendering elastic solutions instable
or untenable in a range of parameters appropriate for
granular materials. (Assuming as above that it is the
stress which retains its form, no $f_{el}$ exists such
that $\sigma_{ij}=\partial f_{el}/\partial u_{ij}$ holds,
because the relevant Maxwell relations are violated.)

Basic to our approach is the assumption that a  finite
elastic region exists. This is universally accepted in
soil mechanics, eloquently supported by de
Gennes~\cite{EdG}, and in fact corroborated by the
stability of sand piles. There are of course well argued
reservations, such as those derived from force arches, or
from the distinction between plastic versus elastic
contacts -- introduced partly due the same desire to
understand yield~\cite{BCC}. However, they are usually
based on an ``intergranular" or microscopic point of
view, and the connection to the elasticity of
macroscopic, continuous media is less than
clear~\cite{gold}. A different argument is the possible
lack of a unique displacement field $U_i$. In our
opinion, elasticity is, at its core, a robust theory: In
spite of crystal defects, frequently rendering $U_i$
ill-defined, elasticity remains valid in solid,
accounting for its capability to sustain shear stresses
-- as long as the defects are stationary. By merely
standing, sand piles demonstrate the same capacity, and
there is no reason why a carefully constructed elastic
theory should not be able to account for all its static,
macroscopic behavior as well. After all, whether a
proposed theory does exactly this should be its ultimate
test.

This ends the introduction. In the following, an
self-contained elastic theory capable of accounting for
mechanical yield and volume dilatancy is developed.
First, we choose an equilibrium state of arbitrary
(thermodynamic) temperature $T$ and packing density
$\rho_c$, with no external forces, especially
gravitation, but with the atmospheric pressure
present~\cite{2}. This virtual state (in spite of its
marginalized stability and individually compressed
grains) is taken as our system of reference, with a
vanishing displacement field $U_i$. The associated free
energy density is $f_1\rho/m$, where $f_1(T)$ is the free
energy, and $m$ the mass, per grain. Turning on the
gravitation and applying external forces will further
strain the granular material and lead to density change.
If the force is small and applied slowly enough, this
change is elastic,
\begin{equation}\label{delta}
\delta\equiv 1-\rho_c/\rho=-u_{nn}.
\end{equation}
As gravitation and normal stress cram the grains, they
lead to finite contact areas between them, and give rise
to finite elastic moduli. The free energy density becomes
\begin{eqnarray}
f=(f_1/m)\rho+\textstyle{\frac 12}
K_bu_{nn}^2+K_au_{k\ell }^0u_{k\ell }^0+\rho Gz,
\label{fe}
\\K_b=\tilde{K_b}\delta^b,\quad
K_a=\tilde{K_a}\delta^a,\qquad\qquad \label{pl}
\end{eqnarray}
with $\tilde{K_a},\tilde{K_b}>0$ for $\delta\geq0$,
$\tilde K_b,\tilde K_a=0$ for $\delta<0$. (The particles
loose contact with one another for $\delta<0$.) Linear
elasticity corresponds to $a,b=0$, while ``Hertz
contacts" imply $a,b=1/2$. However, we did not find a
watertight general reason requiring $a=b$, as assumed in
the {\em quasi-elastic theory}~\cite{EdG}. We believe
this is a question of clarity versus accuracy:
Experiments are better accounted for if $b$ is taken
slightly larger than $a$ (see below). Yet $a$ is the
rather more important exponent, which alone already gives
rise to mechanical yield and volume dilatancy. So $b$
need not be treated with the same care, and one may set
$b=a$ to gain great simplifications in the expressions --
mainly because the Poisson ratio
\begin{equation}\label{poi}
\nu \equiv (3K_b - 2K_a) / (6K_b + 2K_a)
\end{equation}
then lacks critical density dependence. Finally,
Eq~(\ref{pl}) should be taken in the spirit of an
expansion that holds only close to $\rho_c$, or for
$\delta\ll1$.

Given Eqs~(\ref{fe},\ref{pl}) and employing the Eulerian
notation, $2u_{ij}=\nabla _iU_j+\nabla _jU_i-\nabla
_iU_k\nabla _jU_k$, the stress is determined by energy
and momentum conservation (cf.~\cite{tem}) to be $\sigma
_{ij}=(\rho \mu -f)\delta _{ij}-\Psi _{ij}+\Psi
_{ik}u_{kj}+\Psi _{jk}u_{ki}$, where $\Psi _{ij}\equiv
(\partial f/\partial u_{ij})_{\rho ,T}$ and $\mu \equiv
(\partial f/\partial \rho )_{u,T}$. Keeping only the
dominant of the nonlinear terms, we have
\begin{eqnarray}
\sigma _{ij} &=&-K_bu_{nn}\delta _{ij}- 2K_au_{ij}^0
\label{str} \\ &+&\delta^{-1} ({\textstyle\frac
12}bK_bu_{nn}^2+aK_au_{\ell k}^0u_{\ell k}^0)\delta
_{ij}.  \nonumber
\end{eqnarray}
Setting $a,b=0$, only the first line, or the expression
of linear elasticity, remains. The next line comes either
from deriving $K_a,K_b$ with respect to
$\rho=\rho_c(1-\delta)$, leading to a contribution in the
chemical potential $\mu$, or equivalently, with respect
to $u_{nn}=-\delta$, giving rise to additional terms in
$\Psi_{ij}$. Although $a=b\not=0$ in the {\em
quasi-elastic theory}~\cite{EdG}, neither does its stress
contain the second line. As will be shown in a future
work, this is qualitatively alright for silos, because
yield is never a problem here.

To obtain a feeling for the implications of
Eq~(\ref{str}), consider the ``pressure"
$P\equiv\sigma_{kk}/3$, as a function of the compression
$\delta$ and the shear $u_s\equiv\sqrt{u_{\ell
k}^0u_{\ell k}^0}$,
\begin{equation}
P=(1+b/2)\tilde K_b\delta^{1+b} +a\tilde
K_au_s^2/\delta^{1-a}. \label{P}
\end{equation}
$\delta$ is plotted versus $P$ for given $u_s$ and
realistic values of $a, b$, in the left half of Fig.1.
The solid part of the lines are the stable, physical
region, with a positive compressi\-bility; the dashed
lines, showing $P\to\infty$ for $\delta\to0$ are the
unstable, unphysical region. The reason for this can be
understood from the right half of Fig.1, a plot of
$\delta$ versus $u_s$ for given $P$. Starting at the top,
with finite compression and no shear,
$\delta=4\times10^{-4}$, $u_s=0$, we see how the solid
lines decrease, or how volume dilates, for increasing
shear. (Conventional linear elasticity or the
elasto-plastic theory~\cite{plastoEla} yield a straight
horizontal line, stopping at the value asserted by the
Coulomb yield condition.) Shear values right of the
parabola-like curves obviously do not have elastic
solutions, though as mentioned plastic flow solutions of
course must exist.


%
\begin{figure}[tbp]
\begin{center}
\includegraphics[scale=0.5]{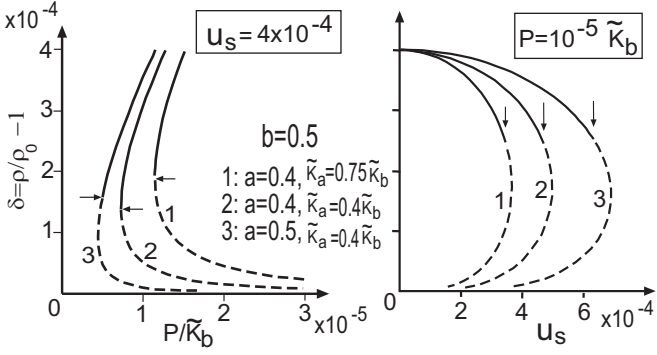}
\end{center}
\caption{Compression $\delta$ versus pressure $P$ (left)
and shear $u_s$ (right). Stability is maintained only
where the lines are solid.}
\end{figure}
Thermodynamic stability, however, is lost even before the
vanishing of the elastic solutions, at the points given
by the arrows, where the solid lines turn into dashed
ones. This is because $f$ of Eq~(\ref{fe}) is convex only
for certain values of $\delta$ and $u_s$: With $f=\frac
12\tilde{K_b}\delta ^{b+2}+\tilde{K_a}\delta
^au_s^2+\cdots$, thermodynamic stability requires
$(\partial ^2f/\partial \delta ^2)(\partial ^2f/\partial
u_s^2)\geq (\partial ^2f/\partial \delta \,\partial
u_s)^2$, or
\begin{equation}
u_s^2/\delta^2\le (2+b)(1+b)K_b/[2a(1+a)K_a].
\label{stability}
\end{equation}
Dashed lines of both figures represent parameters that do
not satisfy this condition. Note that if the medium is
not at all compressed, no finite shear is stable.

For further comparison of this theory to the Coulomb
yield condition (\ref{cy}), we shall in the following
consider three typical experimental setups (see upper
inset of Fig.2): (i)~{\it simple shear test}: an infinite
layer of sand subject to a normal and a shear force
density, $N$ and $S$; (ii)~{\it axisymmetric triaxial
test}: a cylindrical sample of sand subject to a
hydrostatic pressure $p$ and a deviatory normal stress
$q$; (iii)~{\it sand on a slope}: an infinite layer of
sand on a rough, inclined plane with the angle $\theta $.
The implication of Eq~(\ref{cy}) for these experiments
are, respectively~\cite{Nedderman},
\begin{equation}
S/N\leq \tan \varphi _1;\quad q/(2p+q)\leq \sin \varphi
_2; \quad \theta \leq \varphi _1,  \label{3exp}
\end{equation}
with $\varphi _1=\varphi _2$ denoting the Coulomb angle.
Using the above granular elastic theory, setting for
simplicity $a=b$, the respective results may also be
dressed as the same inequalities, though $\varphi_1,
\varphi_2$ are now explicitly given,
\begin{eqnarray}
\tan \varphi _1 &=&\frac{\sqrt{3\left( 1-2\nu \right)
\left( 5a\nu +2\nu -a+2\right) }}{\sqrt{2a}\left( a\nu
-2\nu +a+4\right) }, \label{aos1} \\ \sin \varphi _2
&=&\frac{3\sqrt{1-2\nu }}{2\sqrt{2a(2+a)(1+\nu
)}+\sqrt{1-2\nu }}.  \label{aos2}
\end{eqnarray}
These expressions are easily derived. For the {\em simple
shear test}, the symmetry of the geometry and the force
balance $\nabla _k\sigma_{ik}=0$ (neglecting gravity)
prescribe constant strain, with the displacement given as
$U_x,U_y\sim y$, $U_z=0$. Inserting the nonvanishing
components of the strain, $u_{xy}$ and $u_{yy}=-\delta $,
into Eq~(\ref{str}), we obtain
\begin{eqnarray}\textstyle
\sigma _{yy}=(1+\frac b2)K_b\delta +\frac{4}3(1+\frac
a2)K_a\delta+2a{K_au_{xy}^2}/\delta,\label{stress-i}
\\ \sigma_{xy}=-2K_au_{xy},\qquad \sigma
_{xx}=\sigma_{yy}-2K_a\delta.\label{stress-2}
\end{eqnarray}
The boundary conditions impose $\sigma _{yy}=N$, $\sigma
_{xy}=S$. Solving $\delta$ and $u_{xy}$ as functions of
$N$ and $S$, we find that no solution exists if $N/S$
exceeds a maximal value. In addition, $\delta$ and
$u_{xy}$ must satisfy Eq~(\ref{stability}), which for the
present case is the more stringent one, leading to
Eq~(\ref{aos1}). Note since of Eq~(\ref{stress-i}) is
similar to Eq~(\ref{P}), (they are structurally identical
for $a=b$), it also displays dilatancy.

The strain is again constant for the {\em axisymmetric
triaxial test}. With the displacement vector given as
$U_x\sim x$, $U_y\sim y$, $U_z\sim z$, the strain is
$u_{xx}=u_{yy}=-(\delta +\Delta H/H_0)/2$, $u_{zz}=\Delta
H/H_0$, and $u_{ij}=0$ for $i\neq j$. ($H_0 $ is the
height for $p=q=0$, and $\Delta H$ its change.) Inserting
these into Eq~(\ref{str}) yields the stress tensor,
\begin{eqnarray}\sigma _{xx}=
\sigma _{yy}&=&\textstyle\frac a6q^2/(K_a\delta)
-\frac13q+K_b(1+\frac b2)\delta,\nonumber  \\ \sigma
_{zz}&=&\sigma _{yy}-(\delta +3\Delta
H/H_0)K_a.\label{stress-ii}
\end{eqnarray}
The boundary conditions are: $\sigma _{xx}=\sigma
_{yy}=p$, $\sigma _{zz}=p+q$, $\sigma _{ij}=0$ for $i\neq
j$. Again, $\delta$ and $\Delta H$ do not have solutions
if $q/(2p+q)$ exceeds the value given by Eq~(\ref{aos2}).
[Eq~(\ref{stability}) yields exactly the same constraint
here.]

Due to high external forces, the actual deformations in
both above experiments tend to contain considerable
plastic contributions, invalidating a direct comparison
with the present theory. This is not the case for the
third experiment, {\em sand on a slope}, as its
deformation is due to the comparatively small gravity.
The medium is uniform along the $x$ and $z$ directions
(see inset of Fig.2), so the displacement is of the form
$U_x=U_x(y)$, $U_y=U_y(y)$, $U_z=0$. The stress tensor is
then given by Eq~(\ref{str}) and by integrating $d\sigma
_{yy}/dy=-\rho G\cos \theta $, $d\sigma _{xy}/dy=\rho
G\sin \theta $ under the constraint that the stress
vanishes at the free surface ($y=H$),
\begin{eqnarray}
\sigma _{xx}&=&(1+\textstyle\frac b2)K_b\delta
+\frac23(a-1)K_a\delta +2aK_au_{xy}^2/\delta,\nonumber
\\ \sigma _{xy}&=&-2\mu u_{xy}=-GM(y)\sin
\theta,\nonumber  \\ \sigma _{yy}&=&\sigma
_{xx}+2K_a\delta =GM(y)\cos \theta.\label{stress-iii}
\end{eqnarray}
$M(y)\equiv\int_y^H\rho (y^{\prime })dy^{\prime }$
denotes mass per unit area between $y$ and the free
surface $H$; the total mass $M(y=0)$ is taken as a
constant that does not vary with $\theta$, so both $\rho
$ and $H$ depend on $\theta$. As in the {\em simple shear
test}, the stability condition Eq~(\ref{stability}) is
the more stringent one confining the value of
$\sin\theta$ and leads to Eqs~(\ref{aos1}).

Finally,  we consider what these results imply for the
parameters of the above granular elastic theory,
especially the powers $a$, $b$ and the Poisson ratio
$\nu$. In Fig.2, $\varphi _1$ and $\varphi _2-\varphi _1$
are plotted as functions of $a$ with different values of
$\nu$, rendered in full and dashed lines, respectively.
The angle $\varphi _1$ varies between $15^{\circ }$ and
$35^{\circ }$, with $\left| \varphi _2-\varphi _1\right|
/\varphi _1\lesssim 5\%$ in the parameter regime
$0.2<a<1$ and $0.35<\nu <0.4$. Insisting on $\varphi_2=
\varphi _1=30^{\circ }$ for Coulomb media, we find $a=
0.27$ and $\nu = 0.36$. (The fact that $\left|\varphi_2-
\varphi _1\right|$ diverges for small $a$ indicates that
linear elasticity, $a\rightarrow 0$, is incompatible with
a unique angle of friction.)
\begin{figure}[tbp]
\begin{center}
\includegraphics[scale=0.4]{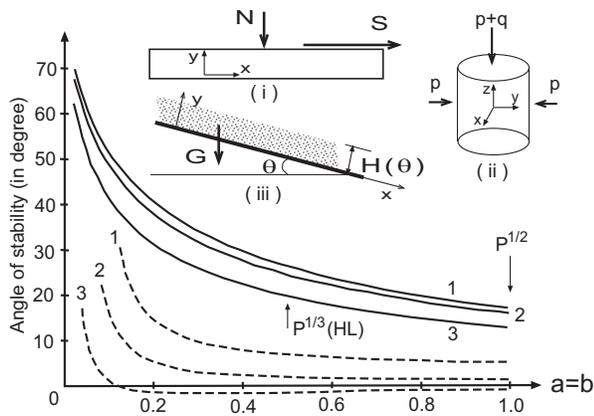}
\end{center}
\caption{Plots of $\varphi _1$ (full lines) and $\varphi
_2-\varphi _1$ (dashed lines) as functions of $a$ for
$\nu=$0.2, 0.3, 0.4 (denoted as 1, 2, 3).}
\end{figure}
Because $\varphi _1$ is the maximum angle for inclined
planes with a layer of sand at rest, we may identify
$\varphi_1$ with {\em Bangold's maximum angle of
stability}. Experimentally, this angle is found to vary
from $30^{\circ }$ (spherical grains) to $60-70^{\circ }$
(angular and rough grains)~\cite{Boltenhagen}. According
to Fig.2, this implies an $a$ between $0$ and $0.4$.

Because we assumed $b=a$, $b$ also varies between $0$ and
$0.4$. With $P\sim \delta ^{1+b}$ or $K_b\sim P^\beta$,
$\beta =b/(1+b)$, cf. Eqs~(\ref{pl}) and (\ref{P}), we
have $\beta$ varying between $0$ and $0.28$ -- smaller
than the $\frac13$ of the Hertz contact~\cite{LL7}. This
is an indication that $a=b$ oversimplifies. Repeating the
above calculation allowing $a\not=b$, we find $\varphi
_1=\varphi _2=30^{\circ }$ by taking $a=0.4 $, $b=0.5$
(ie. $\beta=\frac13$), and $\tilde K_a/\tilde K_b =0.36$.

At higher pressures (from $700$ to $7000$ kg/m$^2$),
$\beta =0.5$ (or $b=1$) was measured and referred to as
the $P^{1/2}$-dependence~\cite{Duffy-Mindlin}). Various
microscopic reasons were proposed for this deviation from
the Hertz law~\cite{Goddard,dG2}. Within the present
framework, this is easily accounted for by the packing
dependence of $b(\rho_c)$, a feature that we shall study
in future works.

It is important to realize that the Poisson ratio $\nu$
as given in Eq~(\ref{poi}) is, in granular materials, not
the same as that given by
\begin{equation}
\nu _{tri}=-\left(\frac{du_{xx}/dq} {du_{zz}/dq}\right)
=\frac 12\frac{b^2+3b+(b^2+3b+6)\nu} {b^2+3b+3+b(b+3)\nu}
\end{equation}
(at $q\to0$), eg  measured in triaxial experiments. [The
second expression is obtained by employing
Eq~(\ref{stress-ii}).] Taking  $b=0.5$, $a=0.4$,
$\tilde{K}_a/\tilde{K}_b=0.36$ as above, and $\delta
=10^{-5}$, we obtain $\nu _{tri}=0.25$, well comparable
to the measured value of around $0.17$ to
$0.25$~\cite{E}. (Enforcing $a=b$ again leads to
discrepancy: Taking $a=b=0.27$ and $\nu$=0.36 as given
above, we have $\nu_{tri}=0.4$.)

\end{document}